\documentclass[preprint,showpacs,amsmath,amssymb]{revtex4}
\usepackage{graphicx}
\usepackage{dcolumn}
\usepackage{bm}

\begin{document}
\title{Magnetotransport in a two-dimensional electron gas\\
in the presence of spin-orbit interaction}
\author{X. F. Wang and P. Vasilopoulos}
\address
{Department of Physics   Concordia University \\ Montreal, QC H3G 1M8, Canada}
\begin{abstract}
We evaluate the transport coefficients of a two-dimensional electron gas (2DEG)
in the presence of a perpendicular magnetic field  and of the  spin-orbit
interaction (SOI) described only by the Rashba term. The SOI mixes the spin-up and spin-down states of
neighboring Landau levels into two new, unequally spaced energy
branches.  The broadened
density of states, as a function of the energy, and the longitudinal resistivity,
as a function of the magnetic field, show
beating patterns in agreement with observations. 
The positions of any two successive nodes
in the beating pattern approximately determine the strength of the  Rashba term.
 A strong SOI
results in a splitting of the magnetoresistance peaks and a doubling of the number of the Hall plateaus.
The peak value in derivative of the Hall resistivity reflects the strength of the SOI.
\end{abstract}
\date{\today}
\maketitle
\section{introduction}

There has been an increasing interest in zero-magnetic-field spin splitting
in one- (1D) and two-dimensional (2D) electron systems due to the spin-orbit
interaction (SOI). Such
systems have potential applications in spin-based transistors
\cite{datt,wang} expected to service in the future quantum computation.
The SOI has been found also important in an unexpected metal-to-insulator
transition in 2D \cite{tutu} hole gas, in spin-resolved ballistic transport \cite{lu},
in Aharonov-Bohm A-B experiments \cite{morp}, and in a spin-galvanic effect \cite{gani}.  
The analysis of the Shubnikov-de Haas (SdH) oscillations in
magnetoresistance measurements has become the main method of measuring
the SOI strength in such systems.

Decades ago theoretical  studies \cite{dres,rash} in 3D semiconductors  found that the spin 
degeneracy should be lifted in inversely asymmetric crystals due to the internal crystal
field. Later, magnetotransport and cyclotron resonance measurements in a 2D hole
system, in a modulation-doped GaAs/AlGaAs heterojunction, showed \cite{stor}
evidence of zero-magnetic-field spin splitting for carriers with finite momentum.
Similar experiments on 2D electron
gases, formed in a GaAs/AlGaAs inversion layer, led to similar conclusions \cite{stei}. 
The first explanation was proposed by Bychkov and Rashba \cite{bych} employing the  
Rashba spin-orbit Hamiltonian, where the spin of finite-momentum
electrons feels a magnetic field perpendicular to the electron momentum in
the inversion plane. Though nonparabolicity of the bulk band structure of
GaAs/AlGaAs could also explain the previous experimental results and bulk
inversion-asymmetry induced spin splitting, at B=0, could dominate in
heterostructures of wide-gap semiconductors, the Rashba SOI  
has been considered the most appropriate reason for the observation of  the
zero-field spin splitting in low-dimensional electron systems, especially
in narrow-gap semiconductors \cite{lomm}. Later, Luo et al. \cite{luo}
investigated the SdH oscillations in a series of GaSb/InAs quantum wells and
concluded that the lifting of the spin degeneracy results from
the inversion asymmetry of the structure which invokes an electric field perpendicular
to the layer. Using the Rashba SOI, they fit the experimental results
and determined the Rashba parameter $\alpha$, which describes the strength of the SOI.
At the same time they concluded that contributions to the SOI from the bulk
$\sim k^3$ term due to a crystal inversion asymmetry are of minor importance.

Generally, the contributions to the spin splitting in the conduction band
of asymmetric heterostructures result from the bulk $\sim k^3$ term due to a crystal inversion asymmetry and from the Rashba $\sim k$ asymmetry. Due to their different momentum dependence, the former  dominates in {\it wide-gap} 
structures with small thickness whereas the later dominates in {\it narrow-gap} structures. It was shown \cite{andr} that the
$k^3$ term leads to anomalous beating patterns while the Rashba term leads to the regular beating patterns in magneto
oscillations. Recently, the well-developed shaping technique in nanostructures
has been used to control the SOI strength in 2D systems of different materials 
\cite{nitt,enge,grun,heid,mole,koga,tsub},
and principally the SdH oscillations are used to measure the Rashba parameter $\alpha$ \cite{engl}.
However,  to our knowledge there are no detailed theoretical treatments of the influence of the SOI on magnetotransport in 2D systems.
We therefore aim at developing a more realistic model
to describe theoretically magnetotransport in  systems with SOI, in which the Rashba term  
dominates, and determine more accurately the parameter $\alpha$.

 In Sec. II we present the
energy spectrum and the density of states (DOS). In Sec. III we present the
results for the transport coefficients and in Sec. IV concluding remarks.
Some auxiliary results are found in the  appendix.

\section{Eigenvectors, eigenvalues, and density of states}

We consider a 2DES in the $(x-y)$ plane and a magnetic field along the $z$
 direction. In   the Landau gauge $\vec{A}=(-By,0,0)$ the one-electron
Hamiltonian including the Rashba term reads

\begin{equation}
H=\frac{({\bf p}+e{\bf A})^{2}}{2m^{\ast }}+\frac{\alpha }{\hbar }%
\left[ {\bf \sigma}\times ({\bf p}+e{\bf A})\right] _{z}+g\mu
_{B}B\sigma _{z},  \label{Ham}
\end{equation}
where ${\bf p}$ is the momentum operator of the electrons, $%
m^{\ast }$ is the effective electron mass,$\ g$ the Zeeman factor, $\mu _{B}$
the Bohr magneton, ${\bf \sigma}=(\sigma _{x},\sigma _{y},\sigma _{z})$ the
Pauli spin matrix, and $\alpha $ the strength of the SOI. 

Using the Landau wavefunctions without SOI as a basis,
we can express the new eigenfunction in the form
($k_{x}$ commutes with the  Hamiltonian (1))
\begin{eqnarray}
\Psi_{k_{x}}({\bf r})
&=&e^{ik_{x}x}\sum_{n,\sigma }\phi _{n}(y-y_{c})C_{n}^{\sigma
}|\sigma \rangle/\sqrt{L_x}  \nonumber \\
&=&e^{ik_{x}x}\sum_{n}\phi _{n}(y-y_{c})
\left(
\begin{array}{c}
C_{n}^{+} \\
C_{n}^{-}
\end{array}
\right)/\sqrt{L_x} ,\ \ \ n=0,1,2,\cdots .  \label{wav}
\end{eqnarray}
Here $\phi _{n}(y-y_{c})=e^{-(y-y_{c})^{2}/(2l_{c}^{2})}
H_{n}( (y-y_{c})/l_{c})/\sqrt{\sqrt{\pi}2^{n}n!l_{c}}$
is the usual harmonic oscillator function, $\omega _{c}=eB/m^{\ast }$
 the cyclotron frequency, $l_{c} =(\hbar
/m^{\ast }\omega _{c})^{1/2}$ the radius of the cyclotron orbit centered
at  $ y_{c}=l_{c}^{2}k_{x}$, $n$ the
Landau-level index, and $|\sigma\rangle$ the electron spin written
as the column vector $|\sigma\rangle =\tiny\left(\begin{array}{c}
1 \\
0
\end{array}
\right) $ if it's pointing up  and $\tiny\left(
\begin{array}{c}
0 \\
1
\end{array}
\right) $ if it's pointing down. Substituting Eq. (\ref{wav})
 in the Schrodinger equation $H\Psi =E\Psi $, multiplying
 both sides by $\phi _{l}(y-y_{c})$,  and integrating over $y$ we obtain
the following system of equations ($E_\pm=\pm
g\mu_{B}B-E$)

\begin{equation}
\left\{
\begin{array}{c}
i(\alpha /l_{c})\sqrt{2l}C_{l-1}^{+}+[(l+1/2)\hbar \omega _{c}
+E_- 
]C_{l}^{-}=0\\
\ \\
\lbrack (l+1/2)\hbar \omega _{c}+E_+ 
]C_{l}^{+}-i(\alpha /l_{c})%
\sqrt{2(l+1)}C_{l+1}^{-}=0
\end{array}
 \right\} , \ \ \ l=0,1,2,\cdots
\end{equation}

This infinite system of equations is solved exactly after decomposing
it into independent one- or two-dimensional secular equations.
Denoting the new subband index by $s$   we obtain

\begin{eqnarray}
[1/2\hbar \omega _{c}+E_-]
C_{s}^{-}&=&0, \ \ s=0;\\
&&\nonumber\\
\left[
\begin{array}{cc}
(s-1/2)\hbar \omega _{c}+E_+
& -i(\alpha /l_{c})\sqrt{2s}\\
i(\alpha /l_{c})\sqrt{2s} & (s+1/2)\hbar \omega _{c}+E_-
\end{array}
\right] \left(
\begin{array}{c}
C_{s-1}^{+}
\\
C_{s}^{-}
\end{array}
\right) &=&0, \ \ \ s=1,2,3,\cdots.
\end{eqnarray}
Corresponding to $s=0$, there is one level, the same as the lowest Landau
level without SOI, with energy $E^+_{0}=E_{0}=1/2\hbar \omega _{c}-g\mu
_{B}B $ and wave function
$\Psi^+_{0}(k_{x})=  (e^{ik_{x}x} /\sqrt{L_{x}})\phi _{0}(y-y_{c})
\tiny\left(
\begin{array}{c}
0 \\
1
\end{array}
\right)
.$
Corresponding to $s=1,2,3,\cdots $, we find two branches of levels with energies  

\begin{equation}
E_{s}^{\pm}=s\hbar \omega _{c}\pm 
[E_{0}^{2}+2s\alpha ^{2}/l_{c}^{2}]^{1/2}.
\label{eig}
\end{equation}
 The
$+$ branch  is described by the wave function

\begin{equation}
\Psi _{s}^{+}(k_{x})=\frac{1}{\sqrt{L_{x}{\cal A}_{s}}}e^{ik_{x}x}\left(
\begin{array}{r}
-i{\cal D}_{s}\phi _{s-1}(y-y_{c})
\ \\
\phi _{s}(y-y_{c})
\end{array}
\right),
\end{equation}
and the $-$ one by

\begin{equation}
\Psi _{s}^{-}(k_{x})=\frac{1}{\sqrt{L_{x}{\cal A}_{s}}}e^{ik_{x}x}\left(
\begin{array}{r}
\phi _{s-1}(y-y_{c})
\ \\
-i{\cal D}_{s}\phi _{s}(y-y_{c})
\end{array}
\right) \text{,}
\end{equation}
where ${\cal A}_{s}=1+{\cal D}_{s}^{2}$ and
\begin{equation}
{\cal D}_{s}=\frac{\sqrt{2s}\alpha /l_{c}}{E_{0} +\sqrt{E_{0}^{2}+2s\alpha
^{2}/l_{c}^{2}}}.
\end{equation}
The density of states (DOS) is defined by
$D(E) =\sum_{sk_x\sigma} \delta(E - E_s^{\sigma})$.
Assuming a Gaussian broadening of width $\Gamma$ we obtain

\begin{equation}
D(E ) =\frac{S_0}{(2\pi)^{3/2}}\sum_{s\sigma}\frac{e^{-(E - E_s^{\sigma})^{2}/2\Gamma ^{2}}}
{l_c^2\Gamma}
\label{dos}
\end{equation}

In Fig. 1 (a) we plot the level energies $E_{s}^{+}$ and $E_{s}^{-}$\
as functions of the level index $s$.
For the case studied
here we have $E_{1}^{-}\simeq E_{0}^{+}$.
Because the level spacing of the $+$  branch is larger than that of
the $-$  branch, the level energy of the $+$
branch increases faster and the  line
through the triangles (not shown) has a  
slope larger than that through the circles (not shown) in Fig. 1 (a).
Here we also notice that
$E_{7}^{-}\simeq (E_{5}^{+}+E_{6}^{+})/2$, $E_{15}^{-}\simeq
E_{13}^{+}$, $E_{25}^{-}\simeq (E_{22}^{+}+E_{23}^{+})/2$.
This difference
in level spacing results directly in the modulation of the
density of states as shown in Fig. 1 (b) and (c). As Fig. 1 (b) shows, where the level
broadening is  small, $\Gamma =0.1$ meV, the DOS as a function of the
energy shows peaks of the same height
except when levels of different branches have the same value and
higher DOS peaks appear. For wider level
broadening, as shown in Fig. 1 (c), with $\Gamma =0.5$ meV, the DOS is
modulated and shows a beating pattern. The
nodes of  this pattern appear when a $-$ branch level is located near
the middle between two $+$ branch levels; thus,
the first node appears near the $E_7^-$ level and the second node near
the $E_{25}^-$ one. The maximum oscillation amplitude appears when
two levels of different branches are degenerate, e.g., at $E_{15}^-\simeq E_{13}^+$ here.

\begin{figure}[tpb]
\includegraphics*[width=80mm,height=70mm]{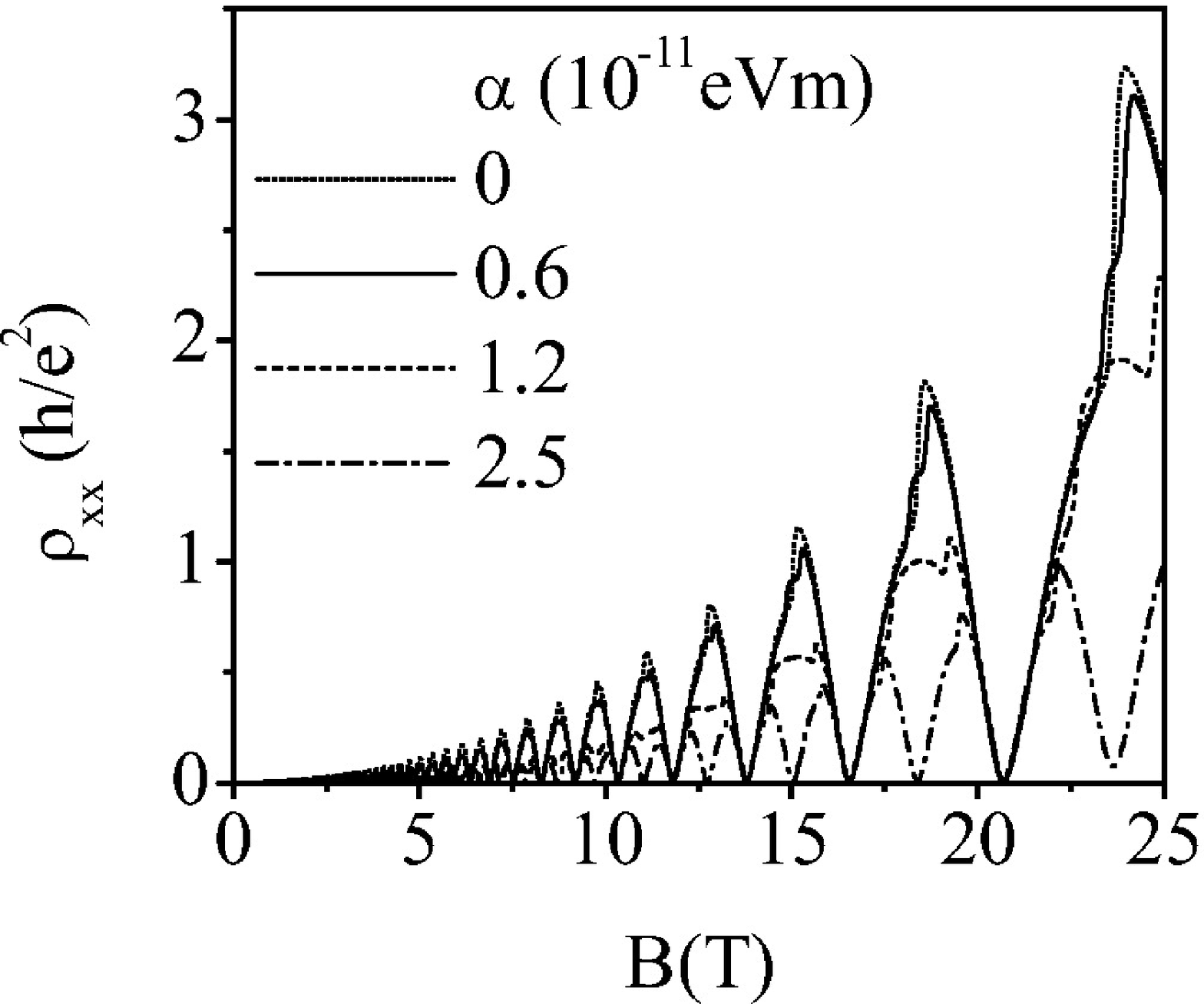}
\caption{(a) Subband energy $E_s$ versus  index s. The triangles are
for the $+$ branch and the circles  for the $-$
branch. (b) Energy (right scale) versus DOS with a subband broadening
$\Gamma=0.1$ meV. (c) The same as in (b) but with
$\Gamma=0.5$ meV. The other parameters are $g=2$, $B=1$ tesla,
$m^*=0.05$, and $\alpha=10^{-11}$eVm.}
\label{fig1}
\end{figure}

\section{Transport coefficients}
\subsection{Analytic results}
For weak electric fields $E_{\nu }$, i.e., for linear responses, and weak
scattering potentials the expressions for the direct current (dc) conductivity
tensor $\sigma_{\mu \nu }$, in the one-electron approximation, reviewed in
Ref. \cite{vas2}, reads $\sigma _{\mu \nu }=$ $\sigma _{\mu \nu }^{d}+\sigma _{\mu \nu
}^{nd}$ with $\mu ,\nu =x,y,z$. The terms $\sigma _{\mu \nu }^{d}$ and $\sigma _{\mu
\nu }^{nd}$\ stem from the diagonal and nondiagonal part of the density
operator $\widehat{\rho }$, respectively,
 in a given basis and $\langle J_{\mu }\rangle =Tr(\widehat{\rho }J_{\mu
})=\sigma _{\mu \nu }E_{\nu }$. In general, we have $\sigma _{\mu \nu }^{d}=$
$\sigma_{\mu \nu }^{dif}+\sigma _{\mu \nu }^{col}$. The term $\sigma_{\mu \nu
}^{dif}$ describes the diffusive motion of electrons
and the term $\sigma _{\mu \nu }^{col}$ collision
contributions or hopping. The former  is given by

\begin{equation}
\sigma _{\mu\nu }^{dif}=\frac{\beta e^{2}}{S_{0}}\sum_{\zeta }
f(E_{s}^{\sigma} )[1-f(E_{s}^{\sigma} )]
\tau ^{\zeta }(E_{s}^{\sigma})
v_{\mu }^{\zeta }v_{\nu }^{\zeta },  \label{diff}
\end{equation}
where $\zeta \equiv (s,\sigma,k_{x})$ denotes the quantum numbers, $
v_{\mu }^{\zeta }=\langle \zeta |v_{\mu }|\zeta \rangle $ is the diagonal
element of the velocity operator $v_{\mu }$, and
$f(\varepsilon )$ the Fermi-Dirac function. Further,
$\tau ^{\zeta }(E_{s}^{\sigma})$ is
the relaxation time for elastic scattering,
$\beta=1/k_BT$, and $S_0$ is the area
of the system.
The term $\sigma _{\mu \nu }^{col}$ can be written  in the form

\begin{equation}
\sigma _{yy}^{col}=\frac{\beta e^{2}}{S_{0}}\sum_{\zeta ,\zeta ^{\prime
}}\int_{-\infty }^{\infty }d\varepsilon \int_{-\infty }^{\infty
}d\varepsilon ^{\prime }\delta \lbrack \varepsilon -E_{s}^{\sigma
}(k_{x})]\delta \lbrack \varepsilon ^{\prime }-E_{s^{\prime }}^{\sigma
^{\prime }}(k_{x}^{\prime })]f(\varepsilon )[1-f(\varepsilon ^{\prime
})]W_{\zeta \zeta ^{\prime }}(\varepsilon ,\varepsilon ^{\prime })(y_{\zeta
}-y_{\zeta ^{\prime }})^{2},  \label{col}
\end{equation}
where $y_{\zeta}=\langle \zeta |y|\zeta \rangle $;  
$W_{\zeta \zeta ^{\prime
}}(\varepsilon ,\varepsilon ^{\prime })$  is the transition rate.
 For  elastic scattering by dilute impurities, of density $N_I$,
 we have

\begin{equation}
W_{\zeta \zeta ^{\prime }}(\varepsilon ,\varepsilon ^{\prime })=\frac{2\pi
N_{I}}{\hbar S_{0}}\sum_{{\bf q}}|U({\bf q})|^{2}
|F_{\zeta \zeta^{\prime }}(u)|^{2}
\delta (\varepsilon -\varepsilon ^{\prime })
\delta_{k_{x},k_{x}^{\prime }-q_{x}},
\label{rat}
\end{equation}
where $u=l_{c}^{2}q^{2}/2$ and $q^2=q_{x}^{2}+q_{y}^{2}$.
$U({\bf q})$ is the
Fourier transform of the screened impurity potential $U({\bf r})=(e^2/4\pi\epsilon_0\epsilon)e^{-k_sr}/r$, where $\epsilon$ is the static dielectric constant, $\epsilon_0$ the dielectric permittivity, 
and $k_s$ the screening wave vector.

\begin{equation}
U({\bf q})
=\frac{e^{2}}{2\epsilon _{0}\epsilon }
\ \frac{1}{(2u/l_{c}^{2}+k_{s}^{2})^{1/2}}
\label{uq}
\end{equation}
In the situation studied here the diffusion contribution given by
Eq. (\ref{diff}) vanishes because
 the diagonal elements of the velocity operator $v_{\mu
}^{\zeta }$ vanish. Neglecting
Landau-level mixing, i. e.,  taking $s^{\prime }=s$,
and noting that $\sigma _{xx}^{col}=\sigma
_{yy}^{col}$, $\sum_{{\bf q}}=(S_{0}/2\pi )\int_{0}^{\infty
}qdq=(S_{0}/2\pi l_{c}^{2})\int_{0}^{\infty }du$, and $\sum_{k_{x}}=
(S_{0}/2\pi l_{c}^{2})$, we obtain
\begin{equation}
\sigma _{yy}^{col}=\frac{N_{I}\beta e^{2}}{2\pi \hbar l_{c}^{2}}%
\sum_{s\sigma }\int_{0}^{\infty }du\int_{-\infty }^{\infty }
d\varepsilon
[\delta
(\varepsilon-E_{s}^{\sigma })]^{2}
f(\varepsilon)[1-f(\varepsilon)]\left| U\left( \sqrt{%
2u/l_{c}^{2}}\right) \right| ^{2}\left| F_{ss}^{\sigma }(u)\right| ^{2}u,
\end{equation}
where
\begin{eqnarray*}
\left| F_{ss}^{-}(u)\right| ^{2} &=&\{L_{s-1}(u)+{\cal D}_{s}^{2}L_{s}(u)
\}^{2}e^{-u}/{\cal A}_{s}^{2}, \\
\left| F_{ss}^{+}(u)\right| ^{2} &=&\{{\cal D}_{s}^{2}L_{s-1}(u)+L_{s}(u)
\}^{2}e^{-u}/{\cal A}_{s}^{2}.
\label{fact}
\end{eqnarray*}

The exponential $e^{-u}$ favors small values of $u$.  Assuming
$b=k_{s}^{2}l_{c}^{2}/2\gg u$ \cite{vas1}, we may neglect the term
$2u/l_{c}^{2}$ in the denominator of Eq. (\ref{uq}) and obtain

\begin{equation}
\sigma _{yy}^{col}=\frac{N_{I}\beta e^{2}}{4\pi \hbar b}\left[ \frac{e^{2}}{
2\epsilon \epsilon _{0}}\right] ^{2}\sum_{s\sigma }\int_{-\infty }^{\infty
}d\varepsilon[\delta (\varepsilon-E_{s}^{\sigma })]^{2}
f(\varepsilon)[1-f(\varepsilon)]I_{s}^{\sigma },
\label{largeb}
\end{equation}
where
\begin{equation}
I_{s}^{\pm }=[(2s\pm 1){\cal D}_{s}^{4}-2s{\cal D}_{s}^{2}+2s\pm 1]/{\cal A}
_{s}^{2}.
\label{in}
\end{equation}

The impurity density $N_{I}$  determines the Landau Level broadening $
\Gamma =W_{\zeta \zeta ^{\prime }}(\varepsilon ,\varepsilon ^{\prime
})/\hbar $. Evaluating $W_{\zeta \zeta ^{\prime }}(\varepsilon ,\varepsilon^{\prime})/\hbar $ 
in the  $u\to 0$ limit without taking into account the SOI,  we obtain $N_{I}\approx 4\pi
[(2\epsilon \epsilon _{0}/e^{2}) ]^{2}\Gamma /\hbar$.

The Hall conductivity $\sigma _{xy}^{nd}$ is given by  

\begin{equation}
\sigma _{xy}^{nd}=\frac{2i\hbar e^{2}}{S_{0}}\sum_{\zeta ,\zeta ^{\prime
}}
f(E_{\zeta } )[1-f(E_{\zeta }^{\prime })]
< \zeta\mid v_x\mid \zeta' >
<\zeta'\mid v_y\mid\zeta >
\frac{1-e^{\beta (E_{\zeta } -E_{\zeta }
^{\prime })}}{(E_{\zeta }-E_{\zeta }^{\prime })^{2}}, \ \ \zeta ^{\prime}\neq\zeta.
\label{hall}
\end{equation}
The evaluation of
Eq. (\ref{hall}) proceeds along the lines of Ref. \cite{vas3} using the
the matrix elements $< \zeta\mid v_\mu\mid \zeta' >,\ \mu=x,y$, given in
the appendix. Taking $\varepsilon _{n+1}^{\sigma }-\varepsilon _{n}^{\sigma
}\approx \hbar \omega _{c}$, leads readily to

\begin{equation}
\sigma _{xy}^{nd}=\frac{e^{2}}{4\pi \hbar }\sum_{s=0}^{\infty
}(s+1)\left[
{\cal B}_{s}\left( f_{s}^{+}-f_{s+1}^{+}\right) +{\cal C}_{s}\left(
f_{s+1}^{-}-f_{s+2}^{-}\right) \right] ,
\label{hall1}
\end{equation}
where

\begin{equation}
{\cal B}_{s}=\frac{1}{{\cal A}_{s}^{2}{\cal A}_{s+1}^{2}}\left\{ \Theta
_{s}^{2}+\frac{2m^{\ast }\alpha {\cal D}_{s+1}}{\hbar ^{2}\omega
_{c}\sqrt{s+1}}\left[ \frac{\alpha {\cal D}_{s+1}}{\hbar \sqrt{s+1}}+(\frac{\hbar
}{m^{\ast }l_{c}}+\Theta _{s})/\sqrt{2}\right] \right\} ,
\end{equation}

\begin{equation}
{\cal C}_{s}=\frac{1}{{\cal A}_{s+1}^{2}{\cal A}_{s+2}^{2}}\left\{ \Theta
_{s}^{\prime 2}+\frac{2m^{\ast }\alpha {\cal D}_{s+1}}{\hbar ^{2}\omega
_{c}
\sqrt{s+1}}\left[ \frac{\alpha {\cal D}_{s+1}}{\hbar \sqrt{s+1}}-(\frac{
\hbar }{m^{\ast }l_{c}}+\Theta _{s}^{\prime })/\sqrt{2}\right] \right\};
\end{equation}
here $\Theta _{s}=1+[s/(s+1)]^{1/2}{\cal D}_{s}{\cal D}_{s+1}$ and
$\Theta_{s}^{\prime }=1+[(s+2)/(s+1)]^{1/2}{\cal D}_{s+1}{\cal D}_{s+2}$. We
notice that if $\alpha =0$, we have $\Theta _{s}= \Theta _{s}^{\prime }=
{\cal A}_{s+1}={\cal A}_{s+2}=1$ and Eq. (\ref{hall1}) becomes  $\sigma _{xy}^{nd}
=(e^{2}/h)\sum_{s=0}^{\infty }f_{s}$, i.e., the conductivity expression
pertinent to the integer quantum Hall. \cite{vas3}

The resistivity tensor, $\rho _{\mu \nu }$ is given in terms of the conductivity tensor.  
We  use the standard expressions $\rho _{xx}=\sigma _{yy}/S$, $\rho_{yy}=\sigma
_{xx}/S$, $\rho _{yx}=\rho _{xy}=-\sigma _{yx}/S$, where $S=\sigma
_{xx}\sigma _{yy}-\sigma _{xy}\sigma _{yx}$.

\begin{figure}[tbp]
\includegraphics*[width=80mm,height=70mm]{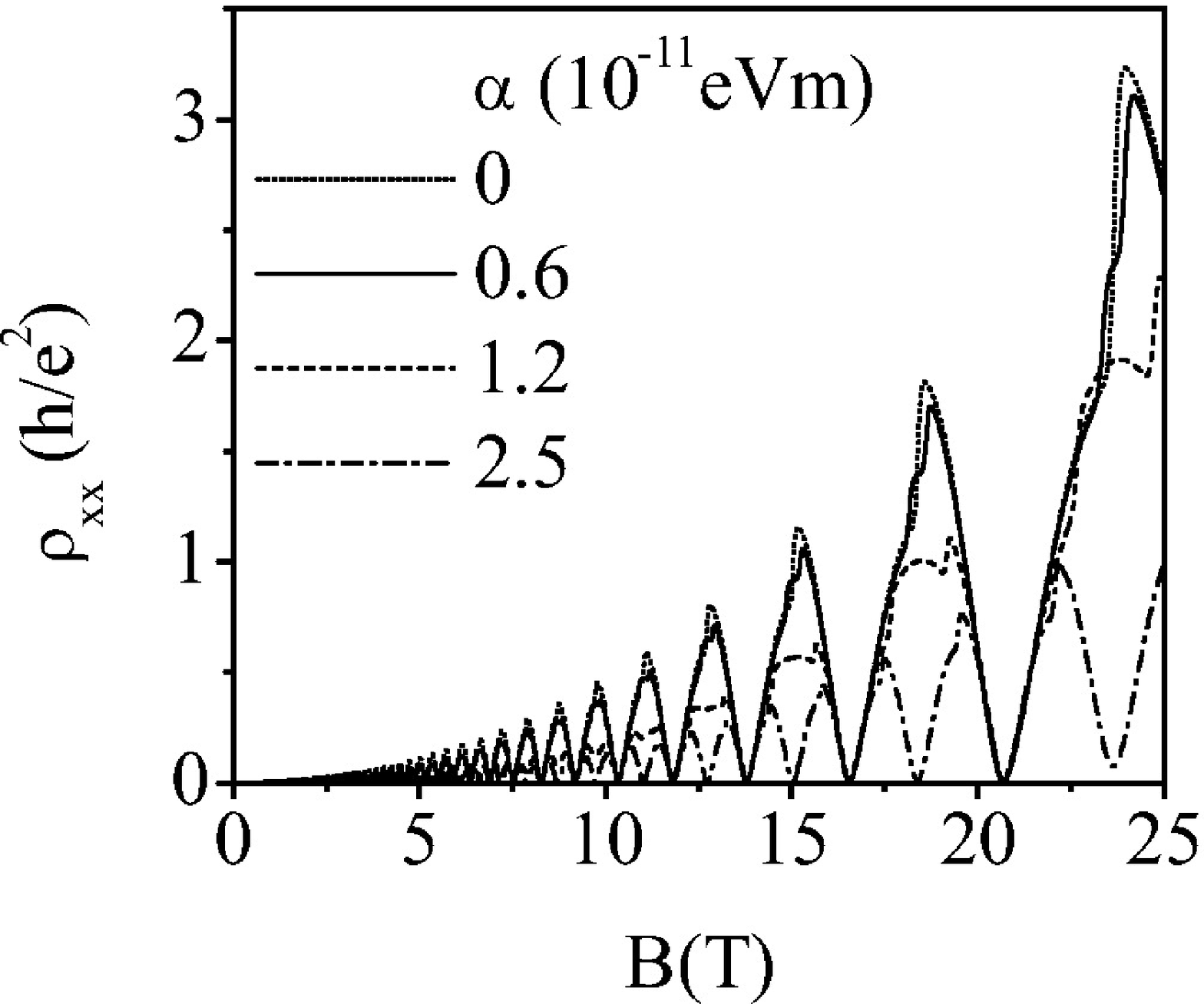}
\caption{ Conductivity $\sigma_{xx}$  as a function of the magnetic field
$B$. The dotted curve is for $\alpha =0$ eVm and the solid one for $\alpha =1.2\times 10^{-11}$ eVm.
The inset shows the oscillations between the fourth and the fifth node.}
\label{fig2}
\end{figure}

\subsection{Numerical results}

In  the numerical evaluation of the conductivity we assume that the $\delta$
functions appearing in Eq. (\ref{largeb})
are broadened and  replace them with the Gaussian function
$(1/\sqrt{2\pi }\Gamma)e^{-x^{2}/(2\Gamma ^{2})}$. Further, if not otherwise
specified, we use the following parameters:
$T=0.4$K,  $\Gamma =1.5$meV, $g=0$, $m^{\ast }=0.05$, $N_{e}=4\times 10^{16}$ m$^{-2}$.
In Fig. 2 we plot   $\sigma_{xx}^{col}$ in the small $u$ limit, cf. Eq. (\ref{largeb}), as     a
function of the magnetic field. The dotted curve shows $\sigma _{xx}^{col}$
in the absence of SOI and the solid one in its presence with $\alpha =1.2\times 10^{-11}$ eVm.
For low magnetic fields and weak $\alpha$ the   conductivity
decreases quickly with $\alpha$  and saturates around
$\alpha=2\times 10^{-11}$eVm. The typical beating pattern appears when the subband broadening is of
the same order as the Landau-level separation. At
high magnetic fields, the effect of SOI is weakened and the beating pattern is
replaced by split conductivity peaks. The
latter approaches that without SOI when the magnetic field becomes very strong.
One of the segments, with a typical beating pattern between the fourth and the
fifth node, is shown in the inset. From Fig. 1 we see that the $m$th node is located near the $n$th
Landau level when $E_{n+m}^{-}\simeq (E_{n-1}^{+}+E_{n}^{+})/2$. Using this expression for large $n$
and small $m$ ($n\gg 1$), we can obtain the ratio of the Landau index over the
magnetic field. 
The result is

\begin{equation}
\frac{n_{m}}{B_{m}}\simeq \frac{(2m-1)(2m+3)e\hbar ^{2}+8gm^{\ast }\hbar
e-4m^{\ast 2}g^{2}\hbar ^{2}e}{32m^{\ast 2}\alpha ^{2}}
\end{equation}
This leads to  

\begin{equation}
\frac{n_{m+1}}{B_{m+1}}-\frac{n_{m}}{B_{m}}\simeq \frac{(m+1)\hbar
^{2}e}{ 4m^{\ast }\alpha ^{2}}.
\end{equation}
If we keep the electron density $N_{e}$ constant and use the definition of the filling factor $\nu=N_e 2\pi l_c^2$, we can  approximate $B_m$ by
 $B_m=\pi \hbar N_{e}/en_m$, and obtain
 
\begin{equation}
\frac{1}{B_{m+1}^{2}}-\frac{1}{B_{m}^{2}}\simeq \frac{(m+1)\hbar e^{2}}{4\pi
m^{\ast 2}\alpha ^{2}N_{e}}
\label{bmod}
\end{equation}
and

\begin{equation}
n_{m+1}^{2}-n_{m}^{2}\simeq \frac{(m+1)\pi \hbar ^{3}}{4m^{\ast 2}\alpha ^{2}}.
\label {amod}
\end{equation}
Eq. (\ref{bmod}) and (\ref{amod}) can be used  to estimate the Rashba parameter $\alpha$.
For instance,  using the inset of Fig. 2 provides $n_{4}=71$ and $n_{5}=87$.
Experimentally, the SdH oscillations in the resistivity of a 2D system, in the presence of SOI,  are usually
viewed as  resulting from a  2D system with two subbands  \cite{tsub,nitt} with the SOI
splitting at the Fermi level $\Delta_R=2\alpha k_F$ serving as the subband separation.
Following this line of reasoning, we can also analyze the results shown in the inset of Fig.2.
The SdH frequency difference between the plus and minus oscillations is
$m\times B_m\simeq 4.8$ Tesla and corresponds to a carrier density
difference $\Delta N=1.16\times 10^15$ m$^{-2}$. This leads [18] to
$\alpha=\hbar k_F \Delta N/(2m^*N_e)=1.1\times 10^{-11}$eVm.
%

%

\begin{figure}[tbp]

\vspace{-0.7 cm}
\includegraphics*[width=80mm,height=70mm]{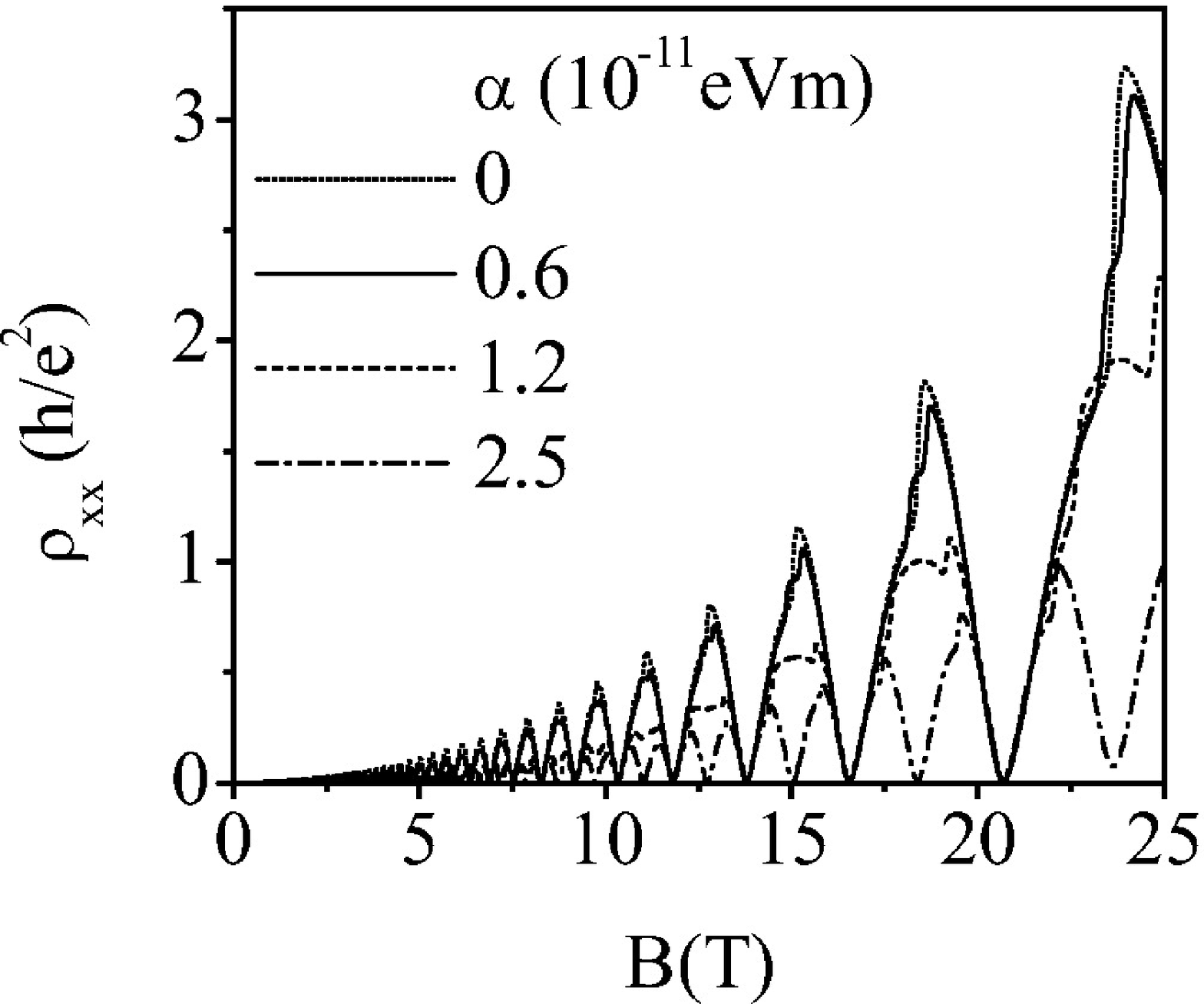}
\caption{The resistivity $\rho_{xx}$  as a function of the magnetic field
$B$ for different values of the parameter $\alpha$ as indicated.}
\label{fig3}
\end{figure}

\begin{figure}[tbp]
\includegraphics*[width=80mm,height=70mm]{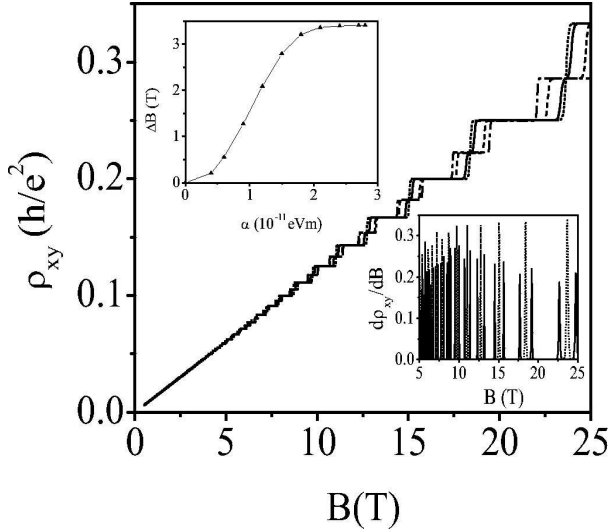}
\caption{Hall resistivity $\rho_{xy}$  as a function of the magnetic field
$B$. The different curves correspond to  different  $\alpha$ and are marked
 as in Fig. 3.  The lower inset shows the derivative of
$\rho_{xy}$ with respect to $B$ versus  $B$.
In the upper inset the  difference $\Delta B$ between the values of the
two peaks in this derivative, into which the $\alpha=0$ peak
near $B=24$ tesla splits, is shown as a function of $\alpha$.}
\label{fig4}
\end{figure}

In Fig. 3 we plot the  resistivity $\rho_{xx}$ for different  strengths
 $\alpha$ as a function of the magnetic field. With the increase of $\alpha$
each resistivity peak becomes lower and gradually splits into two peaks.
However, the shape of the gaps is not affected by SOI. We also notice that all peaks retain almost the same form after splitting.

Figure 4 shows the Hall resistivity $\rho_{xy}$ versus the magnetic field
$B$. For strong magnetic fields we see the
integer quantum Hall effect  plateaus at $h/ne^2$, where $n$ is a integer.
In the presence of  SOI, one more plateau with value $2h/(2n+1)e^2$ appears  between every two
 plateaus of order $n$ and $n+1$.  The size of this new plateau increases with $\alpha$.
It is worth noting that these extra plateaus require rather strong $\alpha$ and 
may easily shrink or disappear if  disorder  is included in the calculation of the Hall resistivity. 
In the lower inset of Fig. 4 we plot the derivative $d\rho_{xy}/dB$
as a function of $B$. Each  peak, corresponding to a sharp jump of the resistivity, splits into two peaks
which separate from each other, by a distance $\Delta B$, with  increasing
$\alpha$. The dependence of $\Delta B$ on $\alpha$ is plotted in the upper inset.  The split increases 
slowly for small $\alpha$ and  saturates at about $\alpha=2\times 10^{-11}$eVm.

\subsection{Comparison with the experiment}

\begin{figure}[tbp]

\includegraphics*[width=80mm,height=50mm]{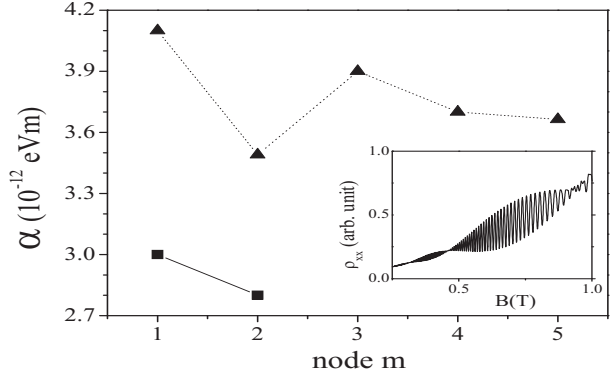}
\caption{ Strength $\alpha$ of the Rashba term, as a function of the observed \cite{das2} node number $m$, extracted from Eq. (\ref{bmod}) and measured node positions $B_m$ and $B_{m+1}$ for sample A (triangles)
and sample C (squares). The inset shows our calculated $\rho$ vs $B$ beating pattern
for sample A. The dotted and solid lines are guides to the eye. }
\label{fig5}
\end{figure}

In the following, we will analyze, using Eq. (\ref{bmod}), two typical measurements of the SdH oscillation in
InGaAs/InAlAs heterostructures, assuming the Rashba term dominates the contribution of the observed zero-field spin splitting.
Ref. [\onlinecite{das2}] provides results for  two samples, A and C. For  sample A, 
with  effective mass $m^*=0.046$ and
sheet density $n_s=1.75\times 10^{12}$ cm$^{-2}$, the positions of the first six nodes are, respectively, at fields, 
$B_1$=0.873T, $B_2$=0.46T, $ B_3$=0.291T, $B_4$=0.227T, $B_5$=0.183T, $B_6$=0.153T. From the positions of any two successive nodes, $B_m$ and
$B_{m+1}$, we extract the Rashba parameter $\alpha$. The results are shown as full triangles in
Fig. (\ref{fig5}); as shown $\alpha$ fluctuates and converges to the average value $\alpha=3.7\times 10^{-12}$eVm
with increasing node number $m$. The consistency of the $\alpha$ values extracted from different nodes convinces us that
the Rashba term is the main cause of the beating pattern here. It may be that bulk SOI  contributes also and results in the variation of   $\alpha$ when different nodes are used. The calculated spin splitting at the Fermi level 
is $\delta_F=2\alpha k_F=2.45$ meV and is the same as the extrapolated result from Ref. [\onlinecite{das2}].
The same analysis has been done for sample C, with $n_s=1.46\times 10^{12}$ cm$^{-2}$ and $B_1$=0.65T, $B_2$=0.312T,
 and $B_3=0.204$T;  the results are shown in Fig. \ref{fig5} as squares. The  zero-field spin splitting at the Fermi level is 1.7meV and is close to the value 1.5meV given in Ref. [\onlinecite{das2}].
In the inset of Fig. (\ref{fig5}) we show the calculated diagonal
resistivity for sample A of Ref. [\onlinecite{das2}] at temperature $T=0.5$ K.
A magnetic-field-dependent subband broadening is adopted,  $\Gamma=\Gamma_0\sqrt{B}$, with $\Gamma_0=0.68$
meV/T$^{1/2}$. The second node is well fitted, while
the first node appears at a slightly higher magnetic field and 40 oscillations are enclosed between them whereas the
number observed in Ref. [\onlinecite{das2}] is 35.

\begin{figure}[tbp]
\includegraphics*[width=80mm,height=50mm]{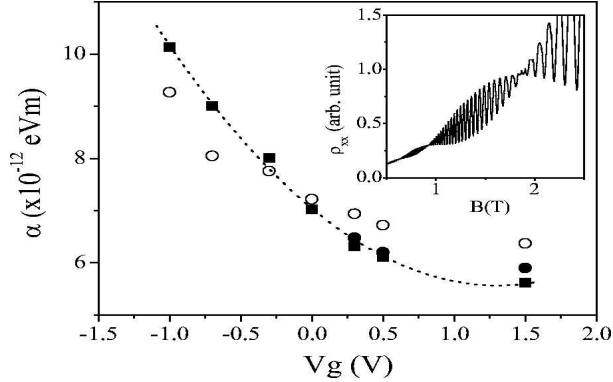}
\caption{Strength $\alpha$ of the Rashba term, as a function of the applied gate voltage,  extracted from Eq. (\ref{bmod}) (squares) and pertinent to the results of
Ref. \cite{nitt}. The latter are shown by the open (solid) circles when the first (second) nodes are fitted
as in Ref. \cite{nitt}. The inset shows our calculated  $\rho$ vs $B$ beating pattern
for $V_g=0.3$V, $T=0.4$K. The dotted curve, produced by $\alpha =7.04-2.26V_g+0.87V_g^2$, is a fit to our results (squares). }
\label{fig6}
\end{figure}

In Ref. [\onlinecite{nitt}] the SdH oscillations of a 2DEG  confined in a gate-controlled InGaAs
layer were observed under
different gate voltages $V_g$, at values -1V,-0.7V, -0.3V, 0V, 0.3V, 0.5V, and 1.5V. The electron effective mass  
ranges from $m^*=0.049$, at $V_g$=-1V,  to $m^*=0.052$,  at $V_g$=1.5V, and the corresponding 
sheet density changes from $n=1.6$ 
 to $n=2.41 \times 10^{12}$ cm$^{-2}$. The first nodes $B_1$ corresponding to the $V_g$ values given above 
are at fields $B_1$= 2.032T, 2.025T,
2.011T, 1.98T, 1.923T, 1.894T, 1.87T, respectively, and the second nodes at  $B_2$= 1.13T, 1.079T, 1.035T, 0.966T, 0.915T,
0.9T, 0.871T. Employing   Eq. (\ref{bmod}) we evaluate the Rashba parameter as a function of the gate voltage. We 
show the results in Fig. \ref{fig6} as filled squares and fit them with the dashed curve, obtained with 
$\alpha =7.04-2.26V_g+0.87V_g^2$. Our results are consistent with those given in Ref. [\onlinecite{nitt}]
 obtained by fitting the first nodes (open circles) of the observed beating pattern with those obtained from an
approximate evaluation of the resistivity; the fitting of the second nodes is shown by the filled circles.
Our calculated magnetoresistivity,  as a function of the magnetic field, for  $V_g=0.3$V and $T=0.4$K,
is shown in the inset of Fig. \ref{fig6}. Here we find the second node is well fitted and a smaller number of
oscillations ($n_2-n_1\simeq 26$) than that observed ($\simeq 28$).
It is worth noting that the value of $\alpha$ we obtained, after analyzing
these two  examples for InAs-based heterostructures,  is of the same order of magnitude as that found by the
microscopic model proposed in Ref. [\onlinecite{andr}] for comparable densities;  from Fig. 3 of Ref. [\onlinecite{andr}]
 the extracted value of $\alpha$ is 
$\simeq 2\times 10^{-11}$ eVm at   
density $n_s=10^{12}$
cm$^{-2}$.

Our  way to extract the  parameter $\alpha$ from the experimental SdH oscillations leads to theoretical results
that are in rather good agreement with those obtained by fitting experimental curves.
One  advantage of Eq. (\ref{bmod}) is that it is independent of the Zeeman splitting. Accordingly, the conclusion can be drawn
that the Rashba effect plays the main role in the formation of beating patterns in the SdH oscillations in
the  measurements discussed above. However, as stated above some mismatches exist, e.g., the $\alpha$ value extracted
from different node sets can vary and not all observed
nodes can be fitted well with the same accuracy. Also, the measured
dependence of the resistivity on the magnetic field is not well recovered by this simple model. This might be a result of the approximations introduced in it,  
e.g., the  neglect of the bulk SOI in the model, the simplified impurity potential, the small $u$ approximation or some unconsidered mechanism influencing the resistivity.     

\section{Concluding remarks}
We studied  magnetotransport in a 2D electron system in the presence of the  Rashba spin-orbit interaction term.
When the subband broadening is much smaller than the Landau level separation,
 the effect of this term on the conductivity is manifested as a splitting of the SdH peaks. For week magnetic fields, with a level broadening  comparable to the Landau
level separation, a beating pattern appears in the conductivity plot as a
function of the magnetic field. By measuring the
position of two successive nodes, we can estimate the  strength $\alpha$ of the
Rashba term. The theory is in reasonably good agreement with the available
experimental observations for $\rho_{xx}$.  
In strong magnetic fields, where  the integer quantum Hall effect is
observed, a sufficiently  strong $\alpha$ creates new plateaus between the integer plateaus in
the Hall resistivity  and splits the  SdH peaks of $\rho_{xx}$.

\section{Acknowledgement}
 We thank Dr. W. Xu for helpful discussions.
 This work was supported by the  Canadian NSERC Grant No.
OGP0121756.

\section{appendix}

The $x$ and $y$ components of the velocity operator read

\begin{eqnarray}
v_{x}=\frac{\partial H}{\partial p_{x}}=\left[
\begin{array}{cc}
-i\hbar \nabla _{x}/m^{\ast }-\omega _{c}y & i\alpha /\hbar\\
&\\
-i\alpha /\hbar & -i\hbar \nabla _{x}/m^{\ast }-\omega _{c}y
\end{array}
\right],
\end{eqnarray}

\begin{eqnarray}
v_{y}=\frac{\partial H}{\partial p_{y}}=\left[
\begin{array}{cc}
-i\hbar \nabla _{y}/m^{\ast } & \alpha /\hbar\\
&                                              \\
\alpha /\hbar & -i\hbar \nabla _{y}/m^{\ast }
\end{array}
\right].
\end{eqnarray}
Setting
${\cal E}_{s}=\omega _{c}l_{c}[s^{1/2} + 
{\cal D}_{s}{\cal D}_{s+1}(s+1)^{1/2}]/\sqrt{2},$
${\cal F}_{s}=\omega _{c}l_{c}[(s-1)^{1/2} + 
{\cal D}
_{s}{\cal D}_{s-1}s^{1/2}]/\sqrt{2},$\\
${\cal G}_{s}=\omega _{c}l_{c} 
\left[ {\cal D}_{s+1}s^{1/2}-{\cal D}_{s}(s+1)^{1/2} -
\sqrt{2}(\alpha/\hbar\omega _{c}l_{c}) {\cal D}_{s}{\cal D}_{s+1}
\right]/\sqrt{2{\cal A}_{s}{\cal A}_{s+1}},$ and
${\cal H}_{s}=\omega _{c}l_{c}  
\left[{\cal D}_{s}s^{1/2}-{\cal D}_{s-1}(s-1)^{1/2} -
\sqrt{2}\alpha/\hbar\omega _{c}l_{c} \right]/\sqrt{2{\cal A}_{s}{\cal
A}_{s-1}},$
we can express the matrix elements of $v_x$ and $v_y$ in the Landau
 representation as follows:

\begin{eqnarray}
\langle \Psi _{s}^{-}(k_{x})|v_{x}|\Psi _{s^{\prime }}^{-}(k_{x}^{\prime
})\rangle =-\frac{\left[ {\cal E}_{s}-\alpha {\cal D}_{s}/\hbar \right]}{
\sqrt{{\cal A}_{s}{\cal A}_{s+1}}} \delta _{s,s^{\prime }-1}\delta
_{k_{x},k_{x}^{\prime }}  
-\frac{\left[{\cal \ F}_{s}-\alpha {\cal D}_{s-1}/\hbar
\right]}{\sqrt{{\cal A}_{s}{\cal A}_{s-1}}} \delta_{s,s^{\prime }+1}\delta
_{k_{x},k_{x}^{\prime }}
\end{eqnarray}

\begin{eqnarray}
\langle \Psi _{s}^{-}(k_{x})|v_{y}|\Psi _{s^{\prime }}^{-}(k_{x}^{\prime
})\rangle =
-\frac{i\left[ {\cal E}_{s}-\alpha {\cal D}_{s}/\hbar \right]}{
\sqrt{{\cal A}_{s}{\cal A}_{s+1}}} \delta _{s,s^{\prime }-1}\delta
_{k_{x},k_{x}^{\prime }}  
+\frac{i\left[ F_{s}-\alpha {\cal D}_{s-1}/\hbar \right]}{\sqrt{
{\cal A}_{s}{\cal A}_{s-1}}} \delta_{s,s^{\prime }+1}
\delta _{k_{x},k_{x}^{\prime }}
\end{eqnarray}

\begin{eqnarray}
\langle \Psi _{s}^{+}(k_{x})|v_{x}|\Psi _{s^{\prime }}^{+}(k_{x}^{\prime
})\rangle = \langle \Psi _{s}^{-}(k_{x})|v_{y}|\Psi _{s^{\prime
}}^{-}(k_{x}^{\prime
})\rangle|_{\alpha\to -\alpha}
\end{eqnarray}

\begin{eqnarray}
\langle \Psi _{s}^{+}(k_{x})|v_{y}|\Psi _{s^{\prime }}^{+}(k_{x}^{\prime
})\rangle =
\langle \Psi _{s}^{-}(k_{x})|v_{y}|\Psi _{s^{\prime }}^{-}(k_{x}^{\prime
})\rangle
|_{\alpha\to -\alpha}
\end{eqnarray}

\begin{eqnarray}
\langle \Psi _{s}^{-}(k_{x})|v_{x}|\Psi _{s^{\prime }}^{+}(k_{x}^{\prime
})\rangle =i{\cal G}_{s}\delta _{s,s^{\prime }-1}\delta
_{k_{x},k_{x}^{\prime }}-i{\cal H}_{s}\delta _{s,s^{\prime }+1}
\delta_{k_{x},k_{x}^{\prime }}
\end{eqnarray}

\begin{eqnarray}
\langle \Psi _{s}^{-}(k_{x})|v_{y}|\Psi _{s^{\prime }}^{+}(k_{x}^{\prime
})\rangle =-{\cal G}_{s}\delta _{s,s^{\prime }-1}\delta
_{k_{x},k_{x}^{\prime }}-{\cal H}_{s}\delta _{s,s^{\prime }+1}\delta
_{k_{x},k_{x}^{\prime }}
\end{eqnarray}

The matrix elements of the  position operator in the $y$ direction
are:
\begin{eqnarray}
\langle \Psi_{s}^{-}(k_{x})|y|\Psi _{s^{\prime }}^{-}(k_{x}^{\prime
})\rangle &=&\frac{l_{c}}{\sqrt{2{\cal A}_{s}{\cal A}_{s+1}}}
\left[ s^{1/2}
+(s+1)^{1/2}){\cal D}_{s}{\cal D}_{s+1}\right]
\delta _{s,s^{\prime }-1}\delta
_{k_{x},k_{x}^{\prime }}\nonumber\\
&+&\frac{l_{c}}{\sqrt{2{\cal A}_{s}{\cal A}_{s-1}}}
\left[(s-1)^{1/2}
+s^{1/2}{\cal D}_{s}{\cal D}_{s-1}\right]
\delta _{s,s^{\prime}+1}
\delta_{k_{x},k_{x}^{\prime }}
+y_{c}\delta _{s,s^{\prime }}
\delta_{k_{x},k_{x}^{\prime }}
\end{eqnarray}

\begin{eqnarray}
\langle \Psi _{s}^{+}(k_{x})|y|\Psi _{s^{\prime }}^{+}(k_{x}^{\prime})\rangle
&=&\frac{l_{c}}{\sqrt{2{\cal A}_{s}{\cal A}_{s+1}}}
\left[{\cal D}_{s}{\cal D}_{s+1}s^{1/2}
+(s+1)^{1/2}\right]
\delta _{s,s^{\prime }-1}\delta
_{k_{x},k_{x}^{\prime }}\nonumber\\
&+&\frac{l_{c}}{\sqrt{2{\cal A}_{s}{\cal A}_{s-1}}}
\left[{\cal D}_{s}{\cal D}_{s-1}(s-1)^{1/2}
+s^{1/2}\right]
\delta _{s,s^{\prime}+1}
\delta_{k_{x},k_{x}^{\prime }}
+y_{c}\delta _{s,s^{\prime }}
\delta_{k_{x},k_{x}^{\prime }}.
\end{eqnarray}
The form factors
 $\left| F_{\zeta\zeta^{\prime}}(u)\right| ^{2}$ read
\begin{eqnarray}
\left| F_{s,k_{x};s^{\prime },k_{x}^{\prime}}^{-}(u)\right| ^{2}
&=&\left|
\langle \Psi _{s}^{-}(k_{x})|e^{i\vec{q}\cdot \vec{r}}|\Psi _{s^{\prime
}}^{-}(k_{x}^{\prime })\rangle \right| ^{2}
\nonumber\\
&=&\left[ (s^{\prime }/
s)^{1/2}L_{s-1}^{s^{\prime }-s}(u) +{\cal D}_{m}{\cal D}_{m^{\prime}}
L_{s}^{s^{
\prime }-s}(u)\right] ^{2} \frac{s!}{s^{\prime }!} \frac{u^{s^{\prime }-s}}{
{\cal A}_{s}{\cal A}_{s^{\prime }}} e^{-u}\delta_{k_{x},k_{x}^{\prime
}+q_{x}},
\end{eqnarray}

\begin{eqnarray}
\left| F_{s,k_{x};s^{\prime },k_{x}^{\prime }}^{+}(u)\right|^{2}
&=&\left|
\langle \Psi _{s}^{+}(k_{x})|e^{i\vec{q}\cdot \vec{r}}
|\Psi _{s^{\prime}}^{+}(k_{x}^{\prime })\rangle \right| ^{2}
\nonumber \\
&=&\left[ (s^{\prime}/
s)^{1/2}{\cal D}_{m}{\cal D}_{m^{\prime }}
L_{s-1}^{s^{\prime}-s}(u)+L_{s}^{s^{\prime }-s}(u)\right] ^{2} \frac{s!}{
s^{\prime }!} \frac{u^{s^{\prime }-s}}{{\cal A}_{s}{\cal A}_{s^{\prime }}}
e^{-u}\delta_{k_{x},k_{x}^{\prime }+q_{x}}.
\end{eqnarray}

\end{document}